%% file: main.tex
\documentclass[sigconf]{acmart}
\makeatletter
\def\@ACM@checkaffil{
    \if@ACM@instpresent\else
    \ClassWarningNoLine{\@classname}{No institution present for an affiliation}%
    \fi
    \if@ACM@citypresent\else
    \ClassWarningNoLine{\@classname}{No city present for an affiliation}%
    \fi
    \if@ACM@countrypresent\else
        \ClassWarningNoLine{\@classname}{No country present for an affiliation}%
    \fi
}
\makeatother

\usepackage{graphicx}
\usepackage{geometry}
\usepackage{makecell}

\AtBeginDocument{%
  \providecommand\BibTeX{{%
    \normalfont B\kern-0.5em{\scshape i\kern-0.25em b}\kern-0.8em\TeX}}}

\copyrightyear{2023} 
\acmYear{2023} 
\setcopyright{acmlicensed}\acmConference[UIST '23]{The 36th Annual ACM Symposium on User Interface Software and Technology}{October 29-November 1, 2023}{San Francisco, CA, USA}
\acmBooktitle{The 36th Annual ACM Symposium on User Interface Software and Technology (UIST '23), October 29-November 1, 2023, San Francisco, CA, USA}
\acmPrice{15.00}
\acmDOI{10.1145/3586183.3606822}
\acmISBN{979-8-4007-0132-0/23/10}

\DeclareUnicodeCharacter{0301}{\'{e}}
\DeclareUnicodeCharacter{0301}{\'{i}}

\newcommand{\sys}{{DiLogics}{}}
\newcommand{\revision}[1]{\textcolor{black}{#1}}

\input{imports}

\begin{document}

\title{\sys{}: Creating Web Automation Programs With Diverse Logics}

\author{Kevin Pu}
\email{jpu@dgp.toronto.edu}
\affiliation{
    \institution{University of Toronto}
}
\author{Jim Yang}
\email{jima.yang@mail.utoronto.ca}
\affiliation{
    \institution{University of Toronto}
}
\author{Angel Yuan}
\email{angel.yuan@mail.utoronto.ca}
\affiliation{
    \institution{University of Toronto}
}
\author{Minyi Ma}
\email{minyi.ma@mail.utoronto.ca}
\affiliation{
    \institution{University of Toronto}
}
\author{Rui Dong}
\email{ruidong@umich.edu}
\affiliation{
    \institution{University of Michigan}
}
\author{Xinyu Wang}
\email{xwangsd@umich.edu}
\affiliation{
    \institution{University of Michigan}
}
\author{Yan Chen}
\email{ych@vt.edu}
\affiliation{
    \institution{Virginia Tech}
}
\author{Tovi Grossman}
\email{tovi@dgp.toronto.edu}
\affiliation{
    \institution{University of Toronto}
}

\renewcommand{\shortauthors}{Kevin Pu, et al.}


\begin{abstract}

Knowledge workers frequently encounter repetitive web data entry tasks, like updating records or placing orders. Web automation increases productivity, but translating tasks to web actions accurately and extending to new specifications is challenging. Existing tools can automate tasks that perform the same logical trace of UI actions (e.g., input text in each field in order), but do not support tasks requiring different executions based on varied input conditions. We present \sys{}, a programming-by-demonstration system that utilizes NLP to assist users in creating web automation programs that handle diverse specifications. 
\sys{} first semantically segments input data to structured task steps. By recording user demonstrations for each step, \sys{} generalizes the web macros to novel but semantically similar task requirements.
Our evaluation showed that non-experts can effectively use \sys{} to create automation programs that fulfill diverse input instructions.
\sys{} provides an efficient, intuitive, and expressive method for developing web automation programs satisfying diverse specifications.

\end{abstract}



\begin{teaserfigure}
  \includegraphics[width=\textwidth]{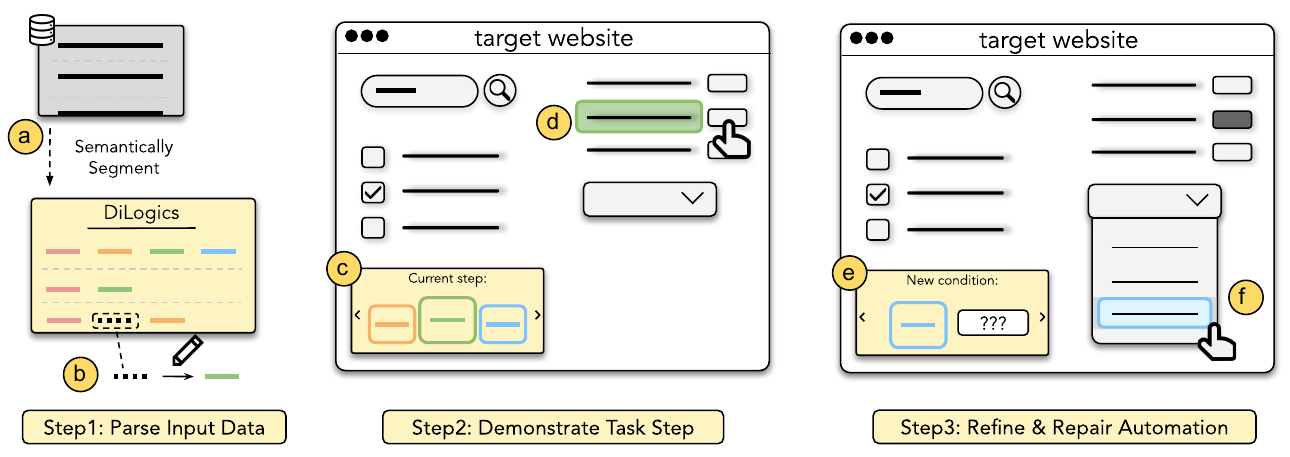}
  \caption{\textbf{The workflow of \sys{}}. (Step 1) To create web automation programs for data entry tasks, users first upload an input file, which is semantically segmented into the steps representing different task specifications \textcircled{\raisebox{-0.5pt}{a}}.
  Users can edit inaccurately segmented steps \textcircled{\raisebox{-0.5pt}{b}}. 
  (Step 2) Users then follow a carousel of current steps \textcircled{\raisebox{-0.5pt}{c}} and demonstrate the corresponding UI actions to fulfill each specification.
  \sys{} highlights semantically relevant web page elements, guiding users to perform demonstrations \textcircled{\raisebox{-0.5pt}{d}}.
  (Step 3) After two iterations of demonstrations, \sys{} learns the mappings between different steps and actions and automates the remaining task steps, generalizing GUI actions based on the specification's semantic meaning. 
  Users can refine the program logics \textcircled{\raisebox{-0.5pt}{e}} at any stage of execution by editing the steps or adding new demonstrations \textcircled{\raisebox{-0.5pt}{f}}.}
  \label{fig:teaser}
\end{teaserfigure}

\maketitle

\section{Introduction}

Interacting with web pages to complete routine data entry and migration tasks is a daily part of many occupations, from receptionists to researchers. But these tasks often require repetitive work that can be time-consuming and unfulfilling. 
Frequently performing these tasks manually can result in human mistakes (e.g., duplicate or missed entries) or frustration ~\cite{leshed2008coscripter}.
In contrast to the manual effort, web automation uses programs to simulate human interaction, creating a faster and more accurate way to complete mundane tasks.
But there is a barrier to creating web automation programs for users without expertise. Through a formative analysis of web automation requests in online platforms such as StackOverflow, we found that non-experts are experiencing difficulty in creating automation programs tailored to their need.
To lower the barriers in the program creation process, existing programming-by-demonstration (PBD) systems, such as SemanticOn and Rousillon~\cite{chasins2018rousillon, pu2022semanticon, sereshkeh2020vasta, dong2022webrobot} allow users to
manually perform a part of the task and construct an automation program based on the demonstrations.

However, while these tools can handle structured repetitive tasks that follow predetermined, uniform program logic, 
they are difficult to generalize when the task contains varied input data that require different page actions to fulfill.
Consider the scenario where an event planner handles employee information from a spreadsheet for booking. They want to enter employee ID into a web form for every colleague traveling to a conference. 
While the data entry step is constant for every employee, inputting ID to the same field (i.e. uniform program logic), the subsequent steps could require different actions targeting different UI elements.
For example, the planner may also need to enter information about dietary restrictions, seating preferences, and planned attendance into multiple data systems and make different selections for every employee (i.e. diverse program logic based on input).
\revision{Another illustrative example is when a coordinator is placing a group lunch order for a social gathering. On the food ordering website, they need to conduct repetitive steps to search for the restaurant, click on the food item, and add it to cart.
Existing tools can automate this uniform logic based on the website structure (e.g. adding each item in the order they appear), but cannot accommodate when requests don't follow such structural order. For example, two requests might order from different restaurants where items are organized differently, and one request requires side dish options while the other notes a dietary restriction.
}
These diverse specifications would likely require the automation program to interact with distinct UI elements in different sequences, \revision{leading to a need for the program to execute a diverse set of logics depending on the input data}.
But the presence and content of these different types of requests might differ for each input. Every local specification (e.g. specify attendance on an event page) might require near identical automation steps, but holistically the different requests are combined and scaled with increasing size of input data to create a hard problem, demanding system intelligence to disambiguate different requests and perform the actions accordingly.
This necessitates the program to be flexible in its choice of execution in order to automate a large variety of steps based on the input data.
To add to the problem, the input data, which describes task requirements, contains enormous heterogeneity in expression (e.g. multiple steps, different phrasing). 
A study on user commands for web actions revealed that people employ various language phenomena, often involving high-level goal description or reasoning \cite{pasupat2018phrasenode}.
While the user could include additional program logic in their automation script to account for diversity in input data and website UI, this extra configuration process can become laborious and error-prone. The resulting program is also task-specific, needs to be maintained, and not scalable.

In this work, we present \sys{}\footnote{\sys{} is an acronym for \textbf{Di}verse  \textbf{Logics}}, a PBD system built upon a program synthesizer~\cite{dong2022webrobot} that assists non-experts in creating web automation programs with diverse and generalizable programming logics.
\revision{The completed program can execute consistent actions for every data input; it also goes beyond symbolic inferences and dynamically executes different UI actions based on the semantic understanding of the task input and web content.}
To create a scalable automation program, \sys{} first semantically segments the input data to decompose the task into more tractable steps (Step 1, Fig.\ref{fig:teaser}). The system represents these steps in a table with a carousel widget (Fig.\ref{fig:teaser}.c) that informs users of the task progress.
Then \sys{} elicits web demonstration for each step (Step 2, Fig.\ref{fig:teaser}), mapping the sequence of UI actions to the description of the step.
At every step, \sys{} leverages natural language processing (NLP) models to scrape web content and locate the most relevant web page content via statistical learning. This way, UI actions (e.g., clicks and selections) are dynamically associated with semantically similar elements on the page. 
In addition, \sys{} employs program synthesis techniques to record the user's actions and their symbolic relationships in the web DOM structure. After a few demonstrations, the system detects the pattern in the action trace and generates an automation program.

As the user demonstrates each step, \sys{} builds a catalog of task steps to UI sequence mappings. Upon entering automation, the system matches each encountered task step to the semantically similar step in the demonstrated catalog and extends the same program logic to fulfill the current condition. When the new step is not meaningfully similar to any previous ones, \sys{} asks the user to demonstrate a new set of UI actions, and adds this step to the catalog for future generalization (Step 3, Fig.\ref{fig:teaser}).
This approach enables flexibility in the execution of the automation program, as it will always employ the most fitting program logic based on semantic similarity, and perform the actions on the relevant element on the current page.
Combining NLP models and program synthesis techniques, \sys{} can generate an automation program that consists of both rule-based structural repetitions, as well as diverse program logics based on the different input data semantics.

We evaluate \sys{}'s usability with 10 participants using four common UI automation data entry tasks. All participants had no prior experience using web automation tools.
We showed that users of \sys{} can successfully create automation programs that satisfy the input requests for every task. 
Despite being novices, participants were able to efficiently construct diverse programming logics by demonstrating different semantic steps. 
Participants also reported that unlike performing manual actions such as copy-and-paste in data entry tasks, mappings task steps to a set of UI actions via demonstration is efficient, generalizable, and reduces mental effort. 
Overall, they found \sys{} intuitive to use and effectively covers diverse scenarios by learning the user demonstrations. 
In the final section, we discuss the implications of \sys{}' design and future works that can adapt our approach to other interactive collaborations between the human and the intelligent system.

This paper contributes the following:

\begin{itemize}
    \item A PBD approach that assists users in creating web automation programs with diverse programming logics.
    \item The technique of semantically categorizing input data and mapping to generalizable UI demonstrations. 
    \item The \sys{} system implementation and user evaluation results assess its effectiveness and usability. 
\end{itemize}

\section{Related Work}
Our work relates to primarily two fields in the PL and HCI communities: web automation and human-AI collaboration. In this section, we identify gaps in existing solutions and draw our design inspirations from these two areas.

\subsection{Web Automation}
The concept of web automation refers to the use of bots to perform tedious and recurring web tasks such as data entry and extraction by simulating human interactions.
It is common for knowledge workers to use web automation in order to accomplish their respective tasks~\cite{li2010here,little2007koala,sugiura1998internet,le2014flashextract}.
For example, data entry workers may need to automate entering data into a digital system for routine tasks such as processing orders or extracting data.


Many tools have been developed to help users to create automation programs. 
For instance, tools like Puppeteer, Selenium, Scrapy, and Beautiful Soup allow developers to select elements and define actions to automate.
Research tools like Sikuli~\cite{yeh2009sikuli} allow users to identify a GUI element (e.g., an icon or a toolbar button) by taking its screenshot.
Using computer vision techniques, it analyzes patterns in the screenshots to locate the appropriate elements when automating GUI interactions. 
Although these tools help lower the effort of creating programs, they all require programming knowledge and cannot disambiguate similar elements or text information.

Even for professional developers, creating automation programs is a non-trivial task.
A study showed that experienced programmers have difficulty writing web macros using common web automation frameworks~\cite{krosnick2021understanding}.
Participants pointed out that a primary hurdle was the labor of checking syntactical element selectors to create their programs, causing inefficiency and errors. 
In addition, the program might not generalize to cross-webpage selections where the elements don't have syntactic similarities.
With our work, users can specify the mappings between a task and its corresponding UI actions via demonstrations.
This saves effort on checking selectors to create a program and enables generalization of UI actions for unseen steps beyond structural similarity.

Alternatively, \revision{researchers have leveraged large datasets of UI to computationally summarize a mobile screen into a coherent natural language phrase \cite{Wang2021Screen2WordsAM}, enabling conversational interaction using large-language-models (LLMs) \cite{Wang2022EnablingCI}, and to ground instructions to UI action sequences \cite{Li2020MappingNL}.} Commercial LLM applications also employ fine-tuned neural models for downstream activity such as UI automation \cite{adept, taxyai}. These tools allow users to prompt the model with high-level natural language intents, which are translated into GUI actions. However, users have limited control outside of describing the task using prompts, and cannot modify the output automation program easily. \sys{} provides a complete pipeline of web automation workflow, from processing input data, to tailoring the program to user demonstration, to refining and handling errors in automation. 

\subsection{Specifying Diverse Programming Logics}


Prior works developed techniques that support users to easily express program logics to satisfy task specifications.
Systems like SemanticOn and PUMICE allow users to specify conditions by demonstrating examples that are (dis)similar to a given specification (e.g., images of two people interacting, weather is hot) \cite{pu2022semanticon, li2019pumice}.
However, they are designed to handle uniform logic -- a binary conditional that determines action or no action applied universally to all content and input. Examples include downloading an image when it contains key objects, or running a macro when the weather is above a certain temperature. In this case, users need to recreate a program when there are multiple specifications that correspond to different UI actions.
Commercial tools like UiPath~\cite{uipath} and iMacros~\cite{iMacro} allow users to set conditional actions on specific page elements via programming, which requires expertise. But, they also lack task understanding to generalize the conditional outside of the symbolic element (i.e. HTML tag) and could not execute different actions based on input data specifications.

Alternatively, researchers designed neurosymbolic languages with both neural and symbolic elements to create programs that satisfy new specifications via approximation.
Neurosymbolic programming is a generalization of classical program synthesis, bridging the gap between deep learning and program synthesis.
Unlike deep learning, neurosymbolic programs can often represent long-horizon, procedural tasks that are difficult to execute using deep networks, and they are also generally easier to interpret and formalize than neural networks~\cite{chen2021web, parisotto2016neuro}.
In contrast to symbolic approaches, neurosymbolic programming does not require all specifications to be hard logical constraints.


However, this approach has been little explored in the context of web automation.
For many years, ML researchers have promoted a ``hybrid model'' that combines the best of both worlds.
As an example, WebQA developed a neurosymbolic system with domain-specific language (DSL) for extracting desired information from datasets that have similar contents but differ in the underlying structures (e.g. DOM structures)~\cite{chen2021web}. 
It omitted, however, user actions during upstream activities (e.g., data collection), limiting it to tasks involving a particular dataset (e.g., data extraction).
For data collection, SemanticOn was able to bridge the communication gap between users' abstract level intent (semantic conditions) and a symbolic system using neural components without defining a DSL~\cite{pu2022semanticon}.
However, while these systems are capable of conditional behavior, their program logic is limited to binary decisions between action and no action, unable to handle diverse specifications with different corresponding actions.
Therefore, current neurosymbolic approaches are either restricted to uniform program logic tasks or require the development of a domain-specific language (DSL) to encode neural model output into symbolic systems (or vice versa), making these approaches not scalable.

\sys{} also employs a neurosymbolic approach for creating programs to automate data entry tasks that involve diverse logics.
Rather than following the same program logic throughout execution, \sys{} learns from user demonstration and uses statistical learning to automate the current step using the most fitting program logic based on task semantics.
By categorizing users' actions into semantic steps, \sys{} learns action patterns and logic-to-demonstration associations using both symbolic inferences and neural network approximation.
The resulting programs extend beyond the uniform program logic that cannot satisfy diverse specifications due to task nature.
Through this construct, \sys{} reduces the level of expertise needed by system designers in other areas to build neurosymbolic programming approaches for their tasks.


\subsection{Programming by Demonstration}

A programming by demonstration (PBD) approach has been adopted by many tools in order to further reduce the expertise required, since users only have to interact with the target applications rather than write code~\cite{lau2003programming,kurlander1993watch,blackwell2000your,mahadevan2022mimic,chen2021umitation}.
Among these application domains are text manipulation~\cite{blackwell2001swyn, lau2000version, mo1992learning, werth1993tourmaline,ni2021recode}, image or video editing~\cite{kurlander1992history,lieberman1994user,maulsby1989metamouse}, and GUI synthesis~\cite{modugno1994pursuit, myers1995garnet,nichols2008mobilization,vaithilingam2019bespoke}.
For web applications, PBD helps build automation programs without requiring users to understand browser internals or reverse-engineer target pages manually. 
CoScripter~\cite{leshed2008coscripter}, Vegemite~\cite{lin2009end}, Rousillon~\cite{chasins2018rousillon}, UiPath~\cite{uipath}, and iMacros~\cite{iMacro} are examples of the PBD approach to web automation.
The resulting programs from user demonstration are represented in visual formats such as a workflow chart \cite{uipath}, a for-loop \cite{chasins2018rousillon}, or in DSL code \cite{iMacro}. These representations require programming expertise, and users have to manually edit program logic which is often nested and convoluted.

Effectively communicating user intent is a major challenge in these PBD systems, and many systems have proposed bridging the gap between user intent and system understanding.
Systems like PLOW~\cite{allen2007plow} and PUMICE~\cite{li2019pumice} allow users to express concepts (e.g., hot weather) in natural language and then learn the concepts to generalize the automation.
\revision{ParamMacros \cite{ParamMacros} allows users to first generalize a concrete natural language question with potential values to identified parameters, and then create a demonstration of how to answer the question on the website of interest.}
Scout~\cite{swearngin2020scout}, Designscape~\cite{o2015designscape}, and Iconate~\cite{zhao2020iconate} allow users to iteratively refine their intent by directly manipulating the AI-generated artifacts.
\revision{SOVITE \cite{sovite} allows users to correct system misunderstanding via direct manipulation of the highlighted entity on the GUI.}
Another work, APPINITE~\cite{li2018appinite}, also encapsulates the user's intent in natural language instructions and clarifies the intention in a back-and-forth conversation with the AI.

Despite promises, specifying intents to cover every case can be tedious.
Furthermore, users may not know all the cases in the first place.
This suggests that tools need to elicit users to better formulate their intent before creating automation programs.
\sys{} addresses this challenge by parsing input data and allowing users to refine their intent continuously during automation by coordinating with our system.


\begin{figure}[H]

\centering
\includegraphics[width=5.5cm]{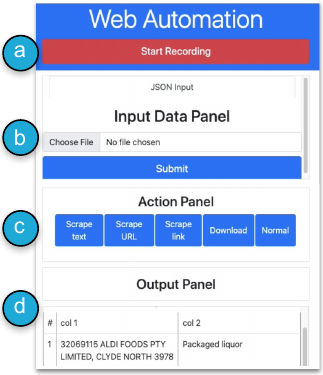}\hspace{1cm}

\caption{A screenshot of the WebRobot system UI.}
\label{fig:webrobot_ui_algorithm}
\end{figure}

\section{Background: WebRobot System}

In this section, we provide necessary information for WebRobot~\cite{dong2022webrobot}, a program synthesizer that constructs a part of the \sys{} system. WebRobot utilizes only web actions and requires no programming expertise, which is consistent with our design goals. 
WebRobot utilizes a no-code approach to synthesize web automation programs based on user demonstration. To create a web automation program for a data entry or scraping task, the user first starts recording their actions (Fig.~\ref{fig:webrobot_ui_algorithm}.a) and optionally uploads a \texttt{JSON} file (Fig.~\ref{fig:webrobot_ui_algorithm}.b) if they need to input data. 
Then, they start demonstrating how to perform the task by choosing an appropriate action type (e.g., Scrape text) in the action panel (Fig.~\ref{fig:webrobot_ui_algorithm}.c) followed by actually performing actions (e.g., clicking the desired text data on the website).
After each scraping action, the output panel displays the extracted data (Fig.~\ref{fig:webrobot_ui_algorithm}.d).
Behind the scenes, WebRobot records every user action with its associated action type. 
At a very high level, WebRobot infers the user intent by generalizing a trace $A$ of user-demonstrated actions to a program $P$ with loops. This generalization is done by ``rerolling'' actions in $A$ into loops in $P$ -- specifically, it infers inner loops first and gradually infers outer loops. 
In particular, $P$ is guaranteed to not only \emph{reproduce} the actions in $A$ but also \emph{generalize} beyond $A$. In other words, $P$ performs more actions after $A$. 
This typically means $P$ is a \emph{loopy program} which ``folds'' actions from $A$ into a loop that can execute for multiple iterations, essentially generalizing user-demonstrated actions based on the same program logic.
Finally, WebRobot executes $P$ to automate the rest of the task, without users manually performing any actions. For more details on the program synthesis algorithm, please refer to the original WebRobot paper \cite{dong2022webrobot}.

\section{Formative Study and Design Goals}
\subsection{Online Web Automation Requests}
To understand the needs and barriers of web automation users, we conducted a formative study analyzing online web automation requests and derive our design goals from the results. 
We collected posts from developer forums like StackOverflow and sub-Reddit communities (e.g. r/automate), as well as commercial tool platforms like UiPath and iMacros forums~\cite{uipathforum, imcarosforum}. 
We used BeautifulSoup4~\cite{beautifulsoup} and the available APIs~\cite{redditapi} to scrape the title, content, comments/replies, and the URL of web posts.
To identify relevant posts and discussions about web automation, we filtered the forums by keywords like ``\textit{UI automation}'', ``\textit{workflow automation}'', and ``\textit{web-scraping}'' and ranked the results by popularity.

After data cleaning, we collected a total of 847 posts. We conducted keyword analysis within post content and identified 53\% of posts as being written by non-experts without web automation or programming expertise, containing phrases like ``\textit{new to [specific tool]}'' or ``\textit{beginner}''. We also found that 61\% of posts were inquiries about how to perform a specific function or approach a task using existing tools. 
This indicates a potential barrier to usage in existing tools as they require specific domain knowledge and experience to utilize.
Combined with a large number of non-expert requests, the learning curve for beginners to accomplish their automation tasks is challenging to overcome.

Our analysis also discovered examples of posts that illustrate conditional specifications. One example is when a user wanted the automation script to iterate through a list of users on a website, and conduct different actions depending on whether the user status is online \cite{imcaros_keyword}.
In another example, the user intended to automate different UI actions based on a text element value \cite{imcaros_specific_text}. Although the element can be easily located by the human, the user expressed difficulty in pragmatically accomplishing this behavior.
In addition, we found that some users desire a simpler way to create and refine web automation programs. For example, one user expressed a need to record web macros and modify them to automate web actions that are generalizable \cite{stackoverflow_record}.


Based on the results of our formative analysis, we identified a barrier for novice users to create web automation programs tailored to their needs. Existing tools require domain knowledge, and cannot fully satisfy conditional automation or generalize the web actions based on the page content.
We also argue that current online requests are limited by the capabilities of existing tools. With higher system intelligence, users could express the need to create more generalizable web automation programs for more complicated tasks that involve conditional steps.
\subsection{Website UI and Content Analysis}
\revision{
We also carried out an informal analysis to identify the common UI action sequences for completing common data entry tasks on websites.
To do so, we analyzed 40 popular websites across 7 genres, including food, shopping, health, entertainment, travel, communication, and scheduling.
These genres are identified by prior study \cite{shi17wob} and extracted from real user requests on forums such as iMacros~\cite{iMacro_forum} and Stack Overflow~\cite{stackoverflow} that discuss the creation of web automation programs for data entry tasks.
We scraped the UI widget types and analyzed the types of UI action sequences needed to complete tasks for these websites. This led us to identify 8 recurring widget categories: buttons with text (appeared in 100.0\% of inspected websites), drop-down menus (77.5\%), checkbox/radio buttons beside a text label (72.5\%), input box for memo or special instructions (27.5\%), calendar widget (20.0\%), plus and minus quantity widget (20.0\%), and seat map (10.0\%).
This investigation helped us determine which are the most common UI widgets that an automation program could encounter. 
We also found that websites utilize consistent GUI elements and interactions for the same categories of functionalities across all pages (e.g. navigation is often associated with buttons or links, search is often associated with an input box). Thus, we design \sys{} to automate GUI tasks under the same web domain where similar UI interactions fulfill the same semantic task. This design scopes the system generalization and assumes that a semantically similar task can be automated via the same UI actions, enhancing the accuracy within the task website domain. 
}


\begin{figure*}[t]
    \centering
    \includegraphics[width=\textwidth]{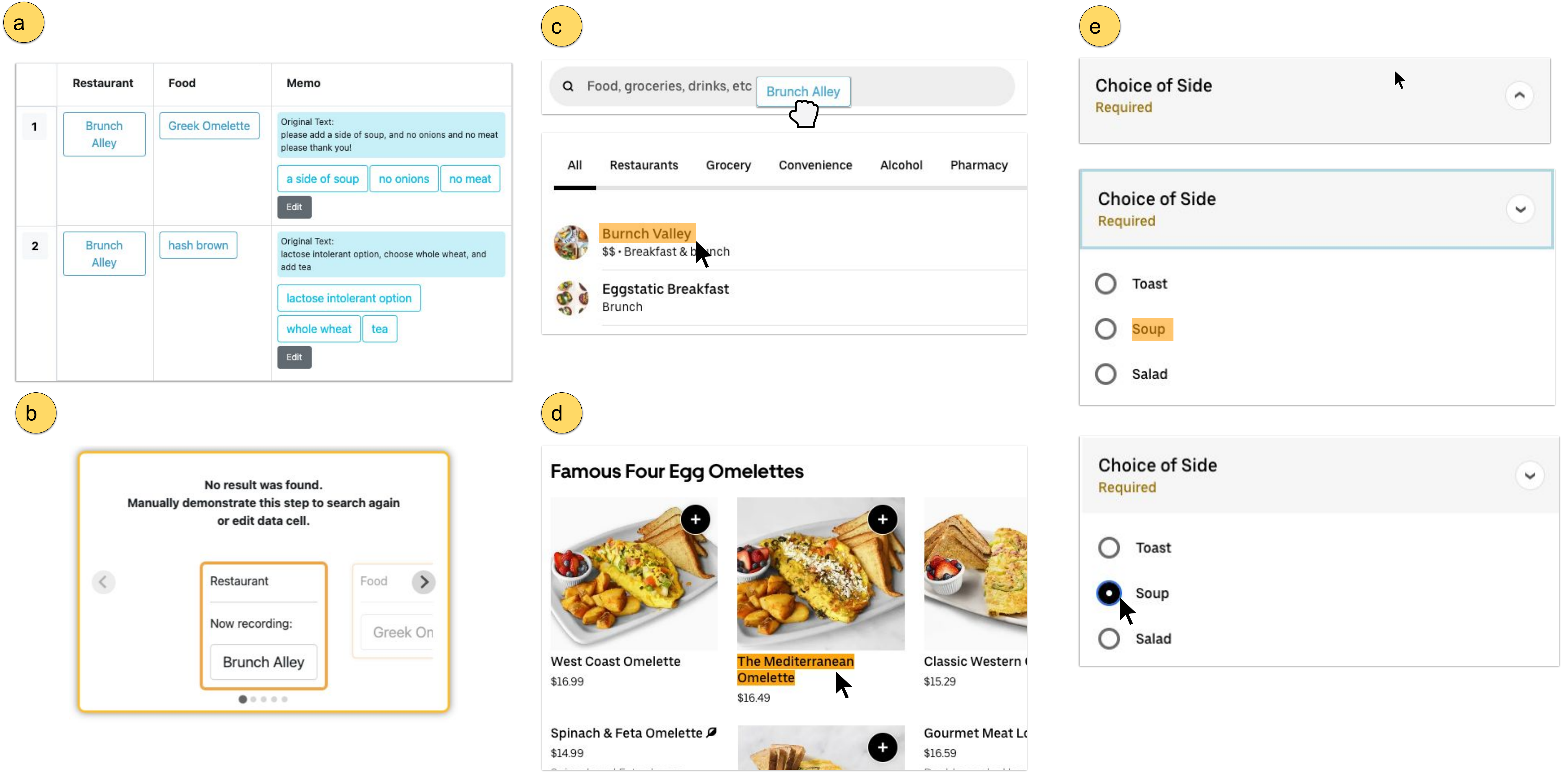}
    \caption{\textbf{An example workflow of \sys{}' user demonstration.} Upon uploading the input file \textcircled{\raisebox{-0.5pt}{a}}, the carousel widget displays all task steps for the current row \textcircled{\raisebox{-0.5pt}{b}}. To demonstrate, the user first drags the restaurant name to the search bar and navigates to the intended page \textcircled{\raisebox{-0.5pt}{c}}. Then, the user moves the carousel to the next slide. \sys{} semantically searches for the dish name and the user clicks on the highlighted result to enter the detail page \textcircled{\raisebox{-0.5pt}{d}}. On this page, the user demonstrates each remaining specification. The first request is adding soup as a side item. \sys{} initially does not find any relevant option, so the user demonstrates by first clicking on the drop-down menu for sides. The system then highlights the relevant option, and the user clicks on the corresponding radio button. The user then moves on to the next specification until the end of the data row.}
    \label{fig:example}
\end{figure*}

\subsection{Design goals}
Based on our formative analyses and prior works, we devised three design goals to help construct our system supporting users in creating web automation programs for data entry tasks with diverse program logics.

\begin{itemize}
    \item \textbf{DG1:} Generalizable specification of the diverse mapping between task steps and user actions.
    \item \textbf{DG2:} 
    Intuitive and natural interaction that constructs automation from demonstration
    \item \textbf{DG3:} Error-handling capability to modify automation and refine step-action mappings accessibly.
\end{itemize}

\begin{figure*}
    \centering
    \includegraphics[width=\textwidth]{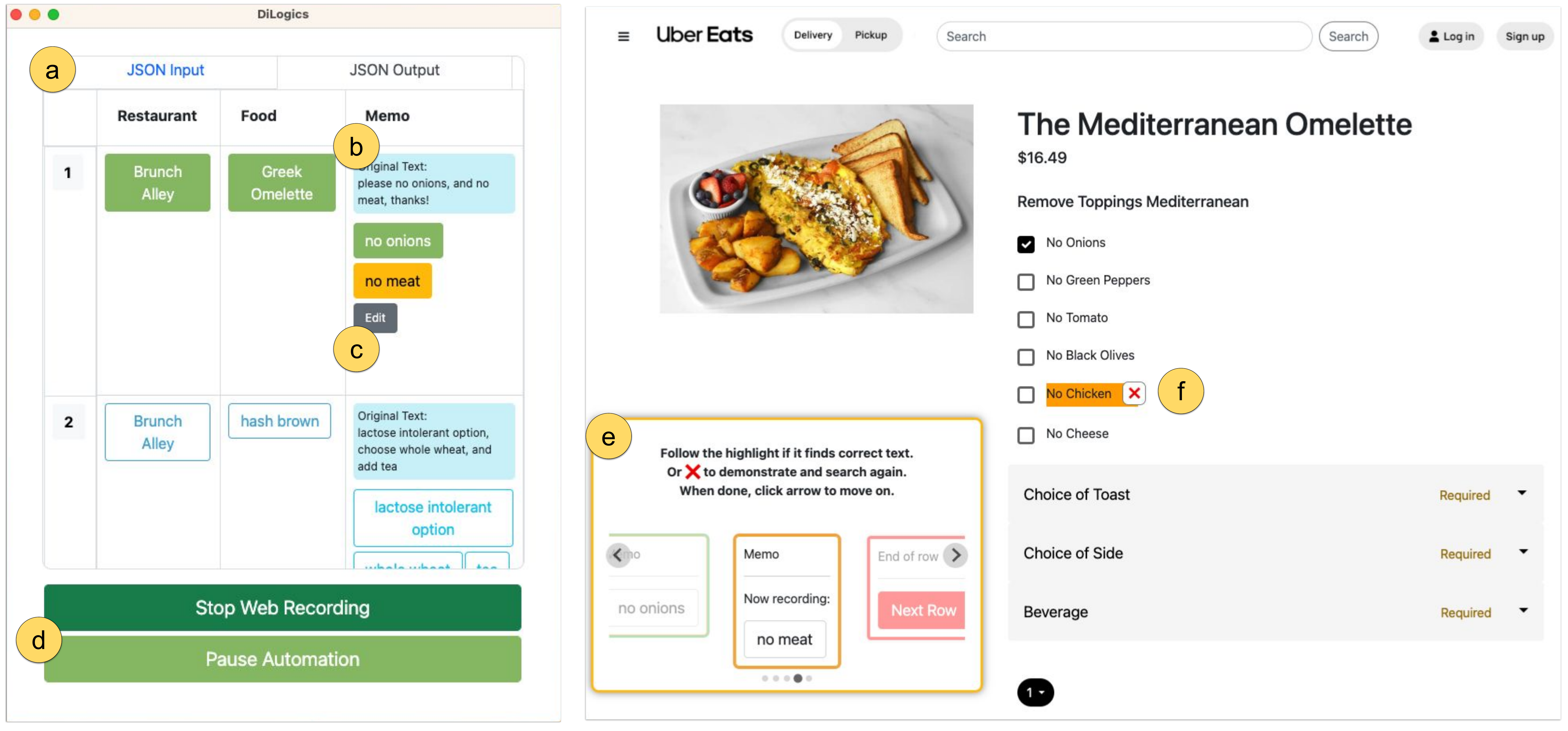}
    \caption{\textbf{\sys{} UI Overview.} Left is \sys{}' extension window. After users upload an input file \textcircled{\raisebox{-0.5pt}{a}}, the data is semantically segmented into task steps \textcircled{\raisebox{-0.5pt}{b}} and rendered into a table. 
    Users can modify the inaccurately segmented step \textcircled{\raisebox{-0.5pt}{c}}. 
    They can also control the flow of the task, and pause automation \textcircled{\raisebox{-0.5pt}{d}}.
    Right is the target website, with an overlaid carousel displaying the progress of the current data row \textcircled{\raisebox{-0.5pt}{e}}. 
    Steps that have been completed are marked green on the data table and carousel, and the current step is marked yellow.
    Users follow the carousel to start the task demonstration. \sys{} semantically searches web page and highlight the most relevant text. 
    If the highlight is incorrect, users can cancel it \textcircled{\raisebox{-0.5pt}{f}}, then edit the step \textcircled{\raisebox{-0.5pt}{c}} or navigate the page to reveal relevant content (e.g. expand the drop-down menu).
    After demonstrating the current step, users can advance to the next slide on the carousel. 
    Users can click ``\textit{Next Row}'' at the end to move on.
    }
    \label{fig:ui}
\end{figure*}

\section{\sys{}}

\subsection{The \sys{} User Experience}
Emma, a corporation clerk, is responsible for processing food orders for all employees at a team-building event. A spreadsheet of everyone's food orders and requests is collected through a survey. Emma could manually enter the selections for each order, but that would be time-consuming and error-prone. Instead, Emma uses \sys{} to efficiently demonstrate common categories of task steps, record her web UI actions, and synthesize an automation program that automatically completes the task for her.
To begin, Emma opens the \sys{} browser extension and uploads the data sheet, displayed as a table with segmented steps for restaurant, dish, ingredient, and dietary restrictions (Fig.\ref{fig:example}.a).
\vspace{-0.5pc}
\subsubsection{Manual Demonstration}
Emma begins the PBD process and moves her attention to the carousel widget on the target web page (Fig.\ref{fig:example}.b), displaying the steps to complete the current food order (i.e. data table row). 
Emma follows the carousel and first searches the restaurant by inputting the name into the search bar and clicking ``\textit{Search}'' (Fig.\ref{fig:example}.c). 
Then, Emma moves to the next carousel step, which is to select the dish. 
\sys{} semantically searches the relevant text content on the page. 
Emma follows the highlight and finds the best fitting dish, then she clicks to navigate to the food details page (Fig.\ref{fig:example}.d).
So far, the demonstrated actions are consistent for every food order, which can be handled by existing PBD tools. But they are limited when every order contains different specifications that require different program logics to fulfill, as we see in later steps.

Again advancing the carousel, the current step is a segmented user request to order ``\textit{a side of soup}'', however, no highlight is shown as the side item menu is not expanded. 
Emma opens the drop-down menu, \sys{} detects a page state change, and highlights the ``\textit{Soup}'' menu item (Fig.\ref{fig:example}.e). 
This sequence of three actions (open menu, search, and click) should not be executed for orders that don't include a side, but existing PDB tools will require configuring a conditional on the symbolic element (i.e. page contains an HTML element with tag ``\textit{Side}'') to handle this case. 
Instead, \sys{} dynamically apply the best fitting program logic by storing this action trace under the category of ``\textit{a side of soup}`` for generalization.
Emma keeps demonstrating each task step following the carousel progression, until arriving at the last slide, which prompts her to complete any remaining action in this row. 
Emma clicks ``\textit{Add To Order}`` to complete this order request and clicks ``\textit{Next Row}`` on the carousel. \sys{} then generates the carousel steps for the next input data row (Fig.\ref{fig:ui}.e).

\subsubsection{Semi-automation}
After Emma demonstrated the second row following the carousel, \sys{} synthesizes an automation program based on user action pattern and enters the semi-automation mode. 
In this stage, the system predicts the next step of action (e.g. ``\textit{Next step is clicking on the highlighted element}'') and prompts the user to review on the carousel widget. Emma can click ``\textit{Confirm}`` to allow \sys{} to automate this step, or click ``\textit{Cancel}'' for incorrect predictions and manually demonstrate the correct step. 
After Emma authorizes the system predictions to fulfill the third-row order, \sys{} enters full automation with the synthesized program.

\subsubsection{Full Automation}
In full automation mode, \sys{} completes the remaining orders row-by-row in the input table, and step-by-step in each row's table cells. 
\sys{} automates actions that are consistent for every order, such as inputting the restaurant name for search or clicking ``\textit{Add to Order}''.
Moreover, \sys{} constructs different program logics to handle different combinations of task steps and generalize to new steps. Emma is pleased to find that \sys{} is able to correctly perform UI actions for ``\textit{add a daily soup}'' even though it is a new condition on a different restaurant page. \sys{} achieves this by semantically matching the step to ``\textit{a side of soup}'', which was previously demonstrated. It can then perform the same UI action sequence but on the ``\textit{daily soup}'' option on the current web page, despite structural differences.

\subsubsection{Refine \& Repair}
Occasionally, when \sys{} encounters a new step that does not semantically match with any previous categories (e.g. ``\textit{select barbeque sauce}''), the system pauses and prompts Emma to demonstrate. 
When \sys{} makes a mistake by highlighting or selecting the wrong element (e.g. highlight ``\textit{Sauce}'' menu heading before the user reveals sauce options in the drop-down), Emma pauses the program to manually cancel the highlight (Fig.\ref{fig:ui}.f), expand the menu, and record new demonstrations to account for new conditions.
This manual effort decreases as \sys{} learns and expands its knowledge of the task semantics. 
Existing PBD tools require users to program the automation for conditional behavior, and could not continuously refine or repair as it executes.
Using \sys{}, Emma did not have to create any conditionals to configure the automation program to handle each task step, \sys{} is able to acquire task understanding and generalize demonstrations based on the step specifications.
Moreover, Emma has the ability to refine program logic and repair errors at any point during the workflow.
Eventually, the program is able to efficiently execute UI actions based on this large datasheet. Emma checks the order list and the shopping cart to verify that the task has been completed, and purchases to confirm the order.

\subsection{\sys{}' Design Rationale and Iteration}
We iteratively designed the \sys{} system based on the feedback from a 10-participant usability evaluation using the prototype. In the initial iteration, \sys{} required users to interact with the extension page to control the flow of demonstration recording, resulting in frequent attention switches between the task website and our tool. Users also needed to manually trigger a semantic search in their action trace which was inefficient. To address these issues, we made significant improvements to the user workflow and interaction process. 
\revision{The final version of \sys{} features a carousel widget (Fig.\ref{fig:ui}.e) overlaid on the target website to guide the task progression, displaying the past, present, and next task steps, following Norman’s visibility principle \cite{norman2013design}. The widget affords and constrains movement back and forth, giving users a sense of direction and task progress.
In addition, by Horvitz’s mixed-initiative UI principles \cite{horwitz_mix-initiative_principle}, we anchor the carousel on the task web page to alleviate effort and reduce context switches.}
In addition, \sys{} now actively searches for semantically relevant page content at every step and website state change, increasing system intelligence to simplify user's workflow.
\revision{As the task goes on, the data table provides color-coded highlighting to signal completion status. This is guided by Norman’s feedback principle \cite{norman2013design} to help users understand \sys{}'s response and confirm their actions.
Moreover, the semi-automation mode after demonstration and before full automation corresponds to Horvitz’s principle of minimizing the cost of poor system guesses \cite{horwitz_mix-initiative_principle}. By walking through one iteration of automation with the user, \sys{} narrows the gulf of evaluation and allows users to validate system actions.}

\begin{figure*}[t]
    \centering
    \includegraphics[width=\textwidth]{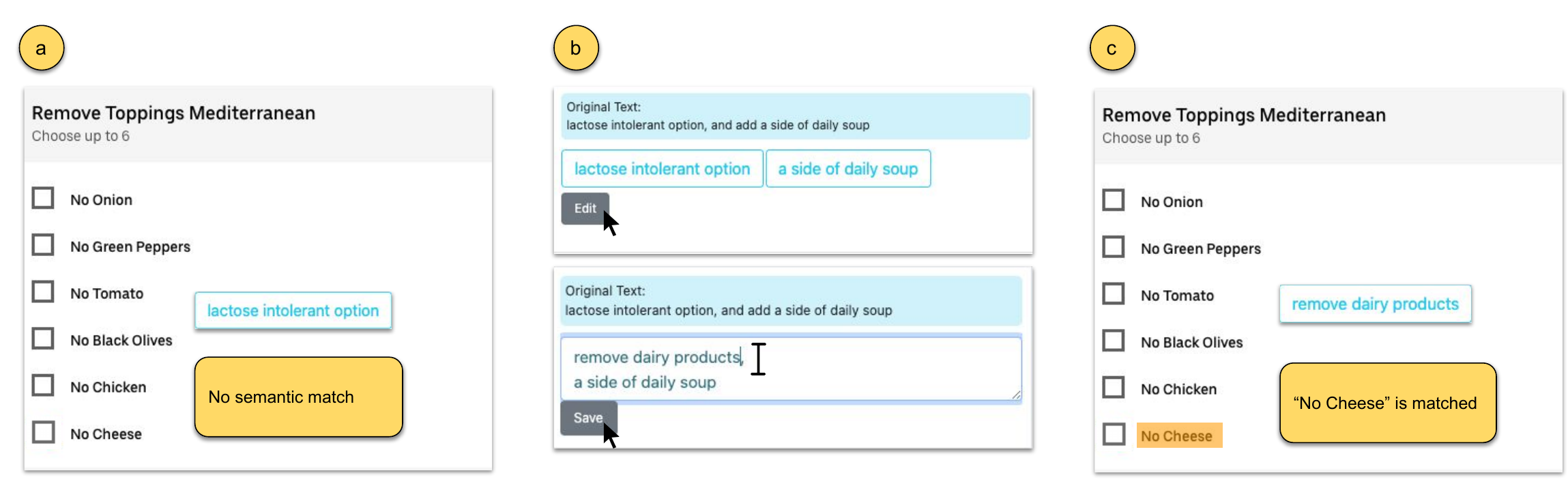}
    \caption{\textbf{Editing data table and specifying step.} During step demonstrations, users might find that a request does not result in any semantic match on the web page content \textcircled{\raisebox{-0.5pt}{a}}. This could be due to incorrect parsing, ambiguity, too high or too low a level of specificity, or limitation of the NLP model. The user can manually edit the itemized step to match the page content more closely \textcircled{\raisebox{-0.5pt}{b}}, and demonstrate the UI actions for this new category of step \textcircled{\raisebox{-0.5pt}{c}}.}
    \label{fig:error_repair}
\end{figure*}

\subsection{Design and Implementation}

We implemented \sys{} as a Chrome browser extension, building upon the core program synthesis engine from WebRobot~\cite{dong2022webrobot}. Primarily, it uses plain \texttt{JavaScript} for recording front-end interactions on the web page. For task step categorization and semantic similarity matching, we adopted two off-the-shelf machine learning models: Sentence-BERT \cite{reimers2019sentence}, a pre-trained network that derives semantically meaningful sentence embeddings that can be compared using cosine-similarity, and SpaCy \cite{matthew_honnibal_2019_3358113}, a trained natural language processing pipeline. The system design and implementation can be separated into three parts, detailed below.

\subsubsection{Step 1: Data Input and Specification Parsing}
To process the input data into tractable steps representing different specifications, users can first upload an \texttt{JSON} input file to \sys{} which renders a data table (Step 1 Fig.\ref{fig:teaser}.a, Fig.\ref{fig:ui}.a,b). While currently only supporting \texttt{JSON}, the input data can be easily extended to other file formats such as CSV, Excel workbook, etc. 
The task file could have inherent structures (i.e. columns and rows of information), but \sys{} further parses the input texts and automatically segments them into semantic steps using SpaCy \cite{matthew_honnibal_2019_3358113}. The data row with the highest number of identified steps will be ranked first. This is to place most of the manual demonstration efforts at the start of the task, allowing users to record actions for most semantic categories upfront, reducing interruption in later automation.
Users can inspect and edit the data if they find incorrectly parsed or ambiguous steps at any point during the demonstration or in automation by pausing the program (Fig.\ref{fig:teaser}.b, Fig.\ref{fig:ui}.c). 

The segmented task steps could also contain specifications that are too abstract or too detailed, which might be misinterpreted by the NLP model and fail to connect to web page content. For example, a user might note that they are ``\textit{lactose intolerant}'' in the specification, but the web page only contains a ``\textit{No cheese}'' option. A semantic search of the original step using yields no match on the page (Fig.\ref{fig:error_repair}.a). The user can then manually rephrase this condition by editing it to ``\textit{remove dairy products}'' (Fig.\ref{fig:error_repair}.b), which \sys{} understands and highlights for demonstration (Fig.\ref{fig:error_repair}.c). Throughout the program creation process, users have the agency to repair and refine task specifications with the help of \sys{}.

\subsubsection{Step 2: Demonstrations and Mapping}
To create an automation program, users start the web recording (Fig.\ref{fig:ui}.d) and demonstrate actions for each task step from the start of the task table.
\sys{}' carousel widget organizes the current row's steps into ordered slides, guiding the users to interact with the website to fulfill the current step as if completing the task manually (Fig.\ref{fig:ui}.e).

Every web macro will be recorded and used to synthesize a repeatable program (e.g. a for-loop). The program synthesis engine based on WebRobot~\cite{dong2022webrobot} enables inferences based on input data and website structures, such as sending each table cell data into a list of input fields in order.
However, it could not generalize the automation to perform different sequences of actions based on the step specification and the page content. 
\sys{} extends WebRobot's functionality by incorporating semantic search in its automation execution. 
After every user action, \sys{} scrapes the web page and highlights the page element that is most semantically relevant to the task step description, as determined by the cosine similarity of the two text phrases~\cite{reimers2019sentence} (e.g. step ``\textit{remove dairy products}'' relates to page option ``\textit{no cheese}''). This intelligent search feature alleviates users' mental effort to process page information.
If the highlight is accurate, users can continue to demonstrate (e.g. click the check box on ``\textit{no cheese}''), or they can cancel the highlight (Fig.\ref{fig:ui}.f) to correct the system by editing the task step or guide the system to highlight the desired region by revealing more relevant page content (e.g. expand a menu to reveal more selections).
\sys{} records this entire sequence of UI actions and maps the task step to the list of macros as a key-value pair. 

As users demonstrate different steps, \sys{} constructs a catalog of step-to-UI action mappings.
In later automation, to perform each task step, \sys{} first inspects the catalog to find the most similar demonstrated step via semantic matching (e.g. new step ``\textit{no meat}'' is similar to ``\textit{remove dairy products}'' in meaning). Then \sys{} generalizes the stored UI action sequence to the current step and automates based on the current highlighted content (e.g. highlight ``\textit{No chicken}'' and click the checkbox next to the option).
Note that since \sys{} constantly searches for and locates the most semantically similar element, the automation can execute macros on the correct UI regardless of website DOM structure, going beyond structural inferences and layout constraints of the task website. 
By recording mappings between diverse specifications and UI actions \sys{} constructs automation programs with malleable programming logic by inserting the proper UI actions for each step in real time to fulfill diverse task specifications.

\subsubsection{Step 3: Automation, Refinement, and Error-handling}
Users follow the progression of the carousel widget to record the manual demonstration for each step in a data table row, which counts as one iteration of the task (e.g. completing one person's food order).
After demonstrating for two iterations (i.e. two rows), \sys{} detects the repetitive pattern in the user action trace and generates an automation program \cite{dong2022webrobot}.
The system then enters a semi-automation stage for the third iteration, where it prompts users with the predicted action for the current step. Users can either confirm to authorize the automation of this step, or cancel in case of incorrect prediction and redo the demonstration for this step.
After confirming the synthesized program's predicted actions in the third row, \sys{} enters full automation.

During full automation, \sys{} iterates the remaining rows of input data and perform corresponding actions based on generalization from the previous demonstration. The carousel advances with the progress of the automation, and each data table cell is marked green when that step is executed. When encountering novel cases that do not match with any steps in the catalog (Fig.\ref{fig:error_repair}.a), \sys{} pauses the automation and elicits a demonstration to fulfill the new specification. After users demonstrate, the system appends a new step-to-macros mapping to the catalog. Through this process, the automation is refined with added categories of demonstration, and the system enhances its capability to handle diverse specifications. 

In the event of a system error during automation (e.g. highlights the wrong element or executes wrong macros), 
users can pause the program (Fig.\ref{fig:ui}.d) to manually inspect and fix the error. They can also edit the data table cell if the step specification is vague (Fig.\ref{fig:ui}.c), or re-record the incorrectly executed step with new demonstrations.
\sys{} provides different error-handling techniques to address input data misinterpretations and system logic errors. Users can gradually transition from manual demonstrations, to evaluating system predictions, and finally to full automation, but always preserve the control to refine and repair the program at every stage.

\section{System Evaluation}

In order to evaluate \sys{}'s general usability in assisting users with diverse program logic data entry tasks across different domains and websites, we conducted an in-person user study. We used the usage
evaluation strategy in the HCI toolkit to guide our study~\cite{ledo2018evaluation}.
The study recruited 10 undergraduate students (6F4M, average age 21.1 y.o., average coding experience 2.3 years, denoted P1-P10) from a large public university.
None of the participants had prior experience with web automation tools.

Since we are implementing a new PBD approach, a within-subject experiment would be difficult as there is no clear baseline to compare to \sys{} in solving automation tasks with diverse specifications.
However, our study reveals findings on the system's usability, coordination with AI, and error-handling in continuous programs, all of which can provide insights into future system designs.



\subsection{Study Design}
Upon signing the consent form, each participant first watched a tutorial video of \sys{}'s interface and features. 
Then participants performed four different task scenarios using \sys{}.
For each task, an input file and a task description were provided.
Each input file contains 10 rows of different requests, and each request requires multiple steps with diverse program logics that need to be accounted for by the participants.
Additionally, the specifications were intentionally designed to have varying levels of abstraction and ambiguity.
This helps examine \sys{}'s refinement features for handling unseen request steps.
The participants could call for the experimenter's assistance at any time during the session. 
After the participants completed the tasks, we conducted a short interview with them regarding their experience. 
Additionally, participants filled out an exit survey with Likert-scale and short-answer questions on system effectiveness, usability, and mental effort \cite{nasa-tlx}.
Each participant was compensated \$25 for their time. 
Each session took 60 minutes and was conducted in person on our machine. All sessions were screen- and audio-recorded. 
Our study is approved by the ethics review board at our organization.

\subsection{Tasks}








To design realistic tasks for users with limited experience with automation tools, we take inspiration from prior studies on common categories of web tasks, UI interactions to accomplish those tasks, and natural language commands to describe the tasks.

Based on QAWob \cite{shi17wob}, a benchmarking study that collected more than 500 website templates and sequences of  GUI actions via crowd-sourcing, we designed our tasks to require common UI interactions such as search, text entry, drop-down, and click.
We also determined our task domains from the common website template categories, such as dining, entertainment, and shopping~\cite{shi17wob}.
Then we constructed four tasks around popular websites in these domains that participants are likely to be familiar with and do repetitive work in, including UberEats, Amazon, GoodRx, and Ticketmaster. Each task involves using \sys{} to create a web automation program that inputs a list of requests to the target website (e.g. food orders with different restaurants and dishes) and fulfills the specifications (e.g. order side, remove ingredient). We limited the number of requests to 10 for each task to standardize the difficulty.

We compose the content of the task input data files following guidance from a dataset of more than 50,000 natural language commands to describe GUI actions on web elements \cite{pasupat2018phrasenode}. The dataset summarizes common categories of language phenomena to express UI action goals in English. The dataset also reveals that many language commands collected from crowd workers go beyond ordinal or visual reasoning (e.g. ``\textit{click the top-most article}'') and use semantic reasoning to describe the goal and target of the GUI actions (e.g. ``\textit{change website language}'' -> a clickable box with text ``\textit{English}'') ~\cite{pasupat2018phrasenode}.
Following these outlines, our data files utilize five commonly classified language phenomena (Relational Reasoning, Substring Match, Paraphrase, Goal Description, and Image Target) \cite{pasupat2018phrasenode} to describe the task specifications. We structured each data file to contain at least 5 diverse steps to maintain the same level of complexity and effort for demonstration. 
\revision{We provide the task input data files so participants can focus on experiencing the full features (e.g. condition failure, new demonstration) of \sys{} within the time constraint of the lab study. Therefore, we didn't let participants specify task input as their instructions may not encompass all \sys{} use cases.} 
Please see the appendix for the specification of each task (Appendix, Fig.\ref{fig:task_table}).

\subsection{Results}


\begin{figure*}[!h]
    \centering
    \includegraphics[width=\linewidth]{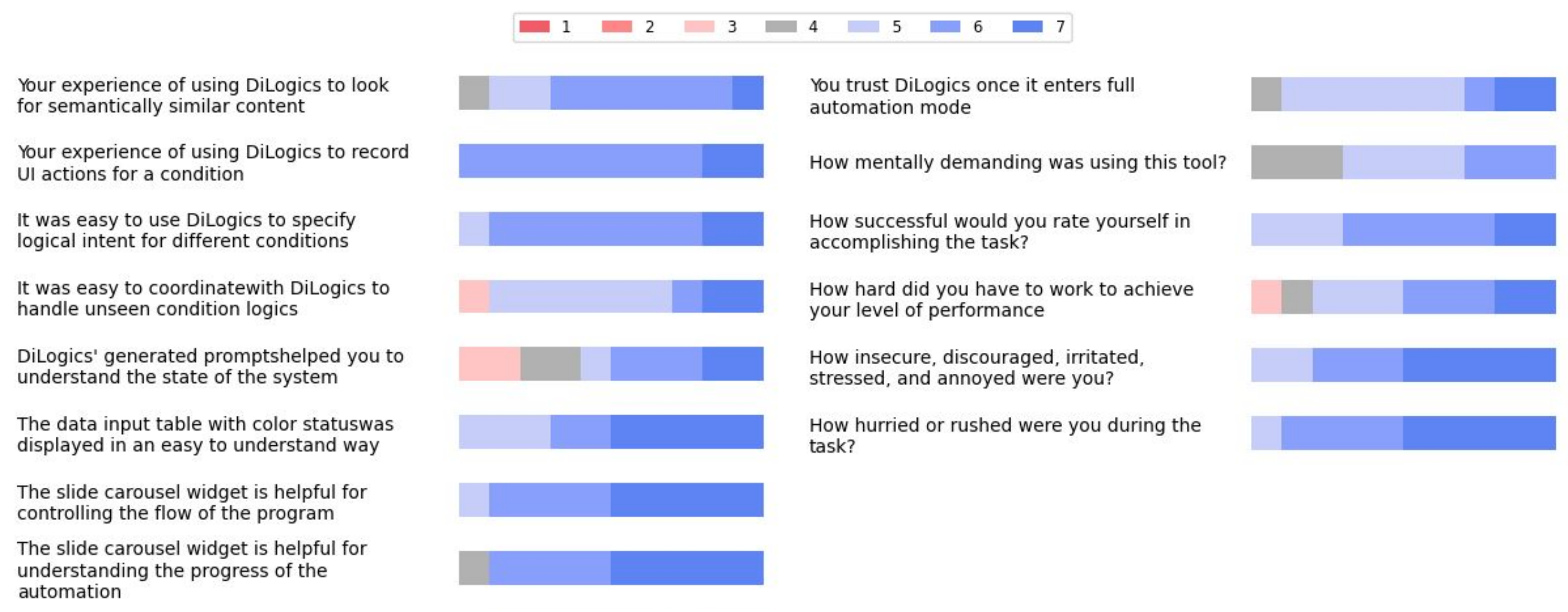}
    \caption{\textbf{Survey Response.} For usability (left) and trust, 1 is very negative, and 7 is very positive. For NASA-Task load index \cite{nasa-tlx} (right), 1 is very high mental demand, effort, stress, feelings of being hurried, and very unsuccessful.}
    \label{fig:survey}
\end{figure*}

\subsubsection{Time and Accuracy}
The user study recorded 40 task completions in total (10 participants x 4 tasks). Table \ref{tab:completiontime} displays analysis for each task, including the average and standard deviation of completion time (in minute:second), task accuracy, and number of attempts to complete this task. 
Task accuracy is based on each row, deemed as one request (e.g. one food order with multiple specifications).
Task accuracy is defined as the percentage of data rows that perfectly satisfied the specification after UI automation (e.g. selecting all the correct options in a food order). 
If users record a demonstration incorrectly, and the system fails to generalize this step in another data row without the user's repair, that data row is counted as incorrect.
The overall average duration to complete a task is 08:01, and the overall average task accuracy is 91.2\%. In 29 of the 40 recorded task completions during the study, participants created automation programs that perfectly satisfied all the specifications in the input files. However, there are cases when users' demonstration fails to generate a program due to human error (e.g. misclick, double click, unfamiliar with task website), reflected in the number of attempts. But errors and retries become less frequent as users learn the tool.

Since our study does not compare \sys{}' approach to a baseline, we keep the task order consistent for every participant and did not conduct with-participant comparisons. However, we do observe participants requiring more attempts and longer time to complete task 1, indicating a learning curve. Some participants also noted that a ``\textit{high level of attention}'' (P9) is required at the start of the study to ``\textit{be careful about the order of the clicks}'' (P7, P9). However, after completing four tasks, participants rated themselves as successful in accomplishing each task on a seven-point Likert scale (\textit{mean=5.9, SD=0.74}, 7 is very successful, Fig.\ref{fig:survey}). Participants also thought they did not need to work very hard to achieve the performance (\textit{mean=5.4, SD=1.26}, 7 is no hard effort at all).
\\
\subsubsection{Effectiveness and Usability}
The participants also rated \sys{}' ease of use and efficiency from ``1 - very negative'' to ``7 - very positive'', detailed in Fig.\ref{fig:survey}. They found their experience positive when using \sys{} to semantically search for content (\textit{mean=5.7, SD=0.82}), recording UI demonstration for a step (\textit{mean=6.2, SD=0.42}), and specifying logical intent for different specifications (\textit{mean=6.1, SD=0.57}).
Participants commented that \sys{} is helpful (P4, P5, P8), interesting (P1, P5, P7), and effective for handling tasks with batch requests (P3, P5, P8). P5 believed that \sys{} is ``\textit{good for repetitive tasks, [where] human might misclick or select wrong content due to large load [of requests]}.''
The workflow from manual demonstration, to semi-automation, and eventually full-automation was also thought to be intuitive, keeping the human in the loop to aid the system's learning process (P1, P4, P5, P7, P8). P4 commented that \sys{} is ``\textit{very intuitive, [with] easy to follow instructions, only takes two trial runs and [\sys{}] knows how to do the rest}.''
Six out of ten participants noted that \sys{} is powerful at information searching and interpretation, automating UI steps across different conditions. 
P9 expressed that ``\textit{semantic matching works for...websites [that] have different layout and structure}.'' 
Overall, participants recognized the system's ability to learn a variety of GUI actions associated with the task specifications and to accurately reproduce desired interactions.

\begin{table}[t]
\sffamily
    \begin{tabular}{l c c c}
    \toprule
    \textbf{Task} & \makecell{\textbf{Time} \\ \textbf{(mm:ss)}} & \makecell{\textbf{Accuracy}} & \makecell{\textbf{\# of} \\ \textbf{Attempts}} \\
    \midrule
    1-UberEats & 08:53 (02:17) & 88.9\% (16.6\%) & 1.8 (0.79) \\
    2-Amazon & 07:32 (02:54) & 100\% (0.00\%) & 1.6 (0.52) \\
    3-GoodRx & 08:33 (02:21) & 94.0\% (8.43\%) & 1.7 (0.48) \\
    4-Ticketmaster & 07:06 (02:33) & 82.0\% (31.6\%) & 1.5 (0.53) \\
    \bottomrule
    \end{tabular}
    \vspace{0.3pc}
    \caption{User study results expressed in average (SD) format. \vspace{-1pc}}
    \label{tab:completiontime}
\end{table}

\subsubsection{Coordination and Error-handling}
Participants found \sys{} relatively easy to coordinate and straightforward, especially for the initial demonstrations to specify program logics (P4, P5, P7). 
Many participants (P2, P3, P4, P5, P8, P9) thought the interaction with the carousel widget provided them with a sense of control (\textit{mean=6.4, SD=0.70}) and helped them understand the progress of the task steps (\textit{mean=6.3, SD=0.95}). 
P6 also noted that the carousel, combined with the semantic highlight, can inform users of the web page content, alleviating the effort of navigating and processing the entire web page.
In addition, once in automation, participants found the execution smooth (P1, P5, P7, P9).
However, during the transitions between user demonstration and automation (i.e. semi-automation, or system pause to demonstrate a new step), half of the participants found themselves sometimes unsure whether it was the users' turn to intervene or the system's turn to automate. Therefore, they desired clearer guidance on the stage of automation and turn-taking.
Participants also commented that sometimes the web macro execution response is not synchronized with the carousel progress and table status coloring, which caused confusion (P1, P3, P5, P6, P7).
This is due to the fact that some UI actions are not grouped into any task step. For example, users might need to click ``\textit{Add to Order}'' at the end of each request, but this implicit action is not categorized in any task description. \sys{} can improve by representing the synthesized program with more context, visualizing past and future predicted actions in addition to the current step.

In terms of error handling, participants utilized \sys{}' error repair techniques at every stage. Nine out of ten users edited the input data table to rephrase segmented steps to the appropriate detail (e.g. ``\textit{lactose intolerant}'' to ``\textit{remove dairy products}''). Seven participants rewound the carousel and re-recorded step demonstration in the event of human error (e.g. misclick or double click).
In addition, eight participants manually repaired UI action errors (e.g. a wrong option is highlighted or selected) and four participants paused during automation to inspect system behavior.
We observed some instances of participants noticing an error but not fixing it.
In the interview, participants expressed that the automation continued (e.g. navigated to a different page) before they could take action to pause and repair. 
P8 and P9 suggested that a redo or undo option in the system workflow would further lower the user's effort to repair errors. 
Future works can make the error-handling interaction more accessible (e.g. more salient and editable execution trace) and provide more processing time or opportunities for users to react to undesired behaviors (e.g. summary of results at the end of task).

\subsubsection{Mental Effort and Trust for AI System}
Overall, participants rated relatively low mental effort (\textit{mean=5.0, SD=0.82}, 7 is not mentally demanding at all) and very low level of stress (\textit{mean=6.3, SD=0.82}) during the study (Fig.\ref{fig:survey}). 
Six participants rated the initial demonstration effort to be high. P7 noted that ``\textit{[demonstration] is a bit heavy as [need] to worry about clicking on something wrong, and to be careful about order the clicks}''. 
As the program shifts to automation, eight out of ten participants reported decreasing mental effort.
However, P5 and P8 believed the semi-automation required the most effort, as the users needed to process and react to the system prompts instead of doing intuitive manual work.

In terms of trust, participants reported a relatively high level of trust when \sys{} starts executing the program in automation (\textit{mean=5.4, SD=0.97}). P4 suggested that ``\textit{would trust the system with more learning [of the tool] and familiarity of the website}'' while P3 mentioned that ``\textit{when the automation seems correct, [they] don't need to watch the system.}''
Participants expressed that the stake of the tasks (P2, P4) and familiarity with the input (P1, P6) are important factors for their trust towards \sys{}.
From the evaluation, \sys{} requires low mental effort after the demonstration phase. And the system elicits a general level of trust from the users. Researchers can focus on providing more guiding feedback and trust cues to lower mental effort and aid users' trust.

    

\section{Discussion}


\subsection{Task to Program Logic Mapping}
To handle diverse program logics, \sys{} creates mappings from each task step's natural language description to its corresponding UI action sequence. This approach encapsulates the UI behaviors inside a natural language label that can be easily compared and generalized. Users define different programming logics for each category of requests. Generalization is built upon the assumption that steps similar in meaning will require similar actions on UI elements with similar affordances. 
From the system evaluation, participants found the task step generalizations interesting (P8), accurate (P7), and even mind-blowing (P1) in terms of capability. P5 notes that the demonstration process is important as ``\textit{[the user] teach[es] the system the rules to automate the steps... helpful to keep human in the loop.}''
\sys{} applies the same semantic intelligence to processing web page content. Once synthesized an automation program, \sys{} generalizes the UI actions to the most relevant UI elements on the current page, despite layout and content differences from the original page user demonstrated on. P1 expressed that they simply needed to ``\textit{let the program search information on the website [and] do series of actions}'', increasing efficiency as users do not need to spend excess time to find the same information.

However, one limitation to the mapping between task steps to UI macros is that the task does not explicitly specify every required action. For example, while users specify their dietary restriction in the food order, they would not specify the need to click on ``\textit{Add to Order}'' when ordering is done, as it is implied. \sys{} captures and automates this type of action with uniform program logic (i.e. does not change based on input) via demonstration. But participants sometimes lose track of the progress as these actions are not represented in the input data table nor the step carousel. Therefore, more signals of system state and current action could be added to improve usability.


\subsection{Neurosymbolic Web Automation}

Web automation tasks often involve the repetition of GUI actions following certain rules on the website DOM or input data structure. 
In symbolic systems, users have a high degree of control to define the rules based on demonstration or programming instructions, if they acquire tool and/or programming expertise \cite{leshed2008coscripter, dong2022webrobot, pu2022semanticon}.
While existing tools can establish symbolic patterns based on UI element properties and web page structures~\cite{iMacro, uipath}, they do not possess the task understanding of the input data nor web page content.
With an increasing volume and diversity of content on the web, symbolic program constructions are limited as the conditions do not apply to the content semantics and do not match the abstraction level of user intent. 
Recent advancement of LLMs leads to a rising need for high-level system understanding to provide an easier method for users to describe their intent in less specific ways that do not require programming expertise.
LLM-powered tools allow intent expression using natural language and can translate abstract intents into executable steps on the web-based on powerful content understanding. 
But current LLM tools offer limited control of the program construction and execution~\cite{adept, taxyai}. Users can only specify intent using prompts and examples, making these tools more similar to an API that responds to individual requests rather than large-scale data.
In addition, pure statistical-learning-based tools can be inconsistent in output generation, but current tools often provide very limited or no error-handling and refinement techniques.

Neurosymbolic systems offer a hybrid model that combines symbolic inferences and similarity-based predictions. 
One main contribution of \sys{} is enabling web automation programs to generalize execution steps based on both symbolic and semantic learning. Mappings are generated between symbolic GUI executions and task semantics to bridge high-level user intent and lower-level web macros.
\revision{Compared to LLM-based web automation tools such as Adept AI, or Taxy AI~\cite{adept, taxyai}, \sys{} provides an end-to-end pipeline from input data parsing to refinement and error repairs, making it a more complete and robust workflow for the downstream tasks of web automation. 
The implementation of \sys{} can adapt to evolving LLM models to harness the power of task and content understanding as the neurosymbolic approach to UI automation is generalizable.}

We argue that this neurosymbolic model can be applied to other automation tasks that involve the semantic understanding of content, such as information organization, content transformation, and generative creation.
Future works can explore how to ground statistical learning models, such as LLMs, in specific task frameworks (i.e. web automation) with general rules (i.e. structural inference on DOM), and how to provide users agency to tailor the process on top of editing prompts.

\subsection{System Scope and Limitations}

The novelty of \sys{}' design is in leveraging semantic understanding to enhance UI automation through a mapping between natural language step categories and web macros. 
This mapping can be established for any web automation tasks where task descriptions can be connected to symbolic UI interactions.
Additionally, the set of interactions for data segmentation, programming logic demonstration, refinement, and error repair can be generalized to any other PBD systems.
While the system is implemented with an existing program synthesizer \cite{dong2022webrobot} and off-the-shelf NLP models \cite{reimers2019sentence, matthew_honnibal_2019_3358113}, \sys{} is not dependent on any specific tool or model.
For example, \sys{} could adopt the latest iteration of LLM to increase the system's semantic understanding capability and content-matching accuracy.

However, the current implementation of \sys{} is limited to understanding text web content and does not support other modalities such as images. This is because the system extracts text-to-element relationships from the web page's HTML to perform semantic search and UI automation. This restriction means that \sys{} cannot infer meaning from images or pure graphical UI (e.g. icons without alt-text), even though human users might express intent in relation to visual information~\cite{pasupat2018phrasenode}.

Despite its generalizability to automate on websites with different DOM structures, \sys{} requires structured input data. The synthesized automation program needs to form a repeatable instruction set (i.e. a loopy program) grounded by symbolic structure on the input (i.e. consistent number of data columns and column ordering). This means that the input data cannot be completely unstructured like a natural language paragraph. In addition, \sys{} requires two iterations of demonstrations to form an automation program; users have to spend some manual effort to perform the first two rows at the beginning of the task.
This means \sys{} cannot perform one-shot automation with a prompt, as can be done by LLM-based automation tools \cite{adept, taxyai}.


Finally, while \sys{}' step to UI actions mapping enables generalizability for similar task semantics, these mappings are restricted to one-to-one relationships. 
\revision{
This was based on the assumption derived from our informal analysis of web UI and content, where for GUI tasks under the same web domain, the same semantic task is fulfilled by similar UI interactions. 
For tasks that span multiple domains and require different UI actions for the same semantic task, \sys{} could lead to high demonstration effort depending on the variety of UI actions and the task size.
The main barrier to one-to-many mappings is the analysis of the webpage state and content.
}
\sys{} is unable to support steps that require updating the system state and repeatedly performing actions to satisfy specifications.
For example, a task step to ``\textit{remove all lactose intolerant option}'' might map to a sequence of actions to search and check ``\textit{no cheese}'', but \sys{} will deem the condition fulfilled after the UI actions are executed. 
This means that the system will not iterate through the web content, identify the state of the condition, and find all applicable options to remove (e.g. also need to select ``\textit{No milk}''), as it requires holistic semantic understanding of the task requirement and the UI states.

As a result of these limitations, we scope \sys{}' capability to handle tasks where the input data is structured, the target website contains text content describing UI elements (e.g. text button, checkbox with text), and the specifications do not require repeated check on website state to fulfill. In the next section, we provide potential directions for future work to overcome these identified hurdles. We also aim to highlight insights in \sys{}' neurosymbolic system design and implementation, as well as the interaction techniques to facilitate continuous human-AI collaboration.

\subsection{Future Work}
Based on user feedback from the system evaluation, we provide several directions to aid future system designs.
First, as discussed in \sys{}' limitations, the semantic task understanding is limited to textual web page content. Future works can leverage multi-modal neural networks, such as CLIP \cite{radford2021clip}, to understand the website's visual content as well, expanding the capability to handle user intents that involve visual reasoning (e.g. click on the dish with fish on the image).
A parallel approach is to construct a knowledge graph of the website content, which could connect different contents in the form of text, image, video, and/or audio \cite{Li_2021_screen2vec}. This can generate a holistic view of the web page and even relate different pages, increasing information searching capabilities during the automation. 
Future works could potentially achieve this using content summarization techniques to transform and embed all types of content in the same space and compare similarities.
\revision{The additional high-level understanding of the web page content and the task goal can potentially expand the existing one-to-one task-UI mapping to a one-to-many relationship, enhancing generalizability.}

Another limitation of \sys{} is that the automation can not perform logics that require constant analysis of the website state (e.g. check if all lactose intolerant options are chosen). Future systems can constantly analyze the website's state and task completion status after every UI action or DOM change. This requires storing the state of the website and understanding the status of UI elements at a semantic level (e.g. the ``\textit{No cheese}'' option is selected, but the condition is not fulfilled as ``\textit{No milk}'' is not selected).


Finally, to further reduce user effort and make web automation programs easier and more accessible to create, future works can potentially derive patterns of execution based on previous task completion. Since every task generates an automation program, there might exist many overlaps in steps and programming logics for tasks in the same domain. Researchers can leverage the history of these completed tasks to make predictions on a new task. If the task can be processed through past similar tasks, the user might not even need to perform initial demonstrations to start program generation; the past tasks may already be capable of predicting the execution of the current one.

\section{Conclusion}
To support creating web automation with diverse specifications, we designed and developed \sys{}, a PBD tool that assists users in segmenting task requests and synthesizes programs based on user demonstration of example steps. The steps are mapped to sequences of UI actions and can be generalized using both symbolic inferences and semantic similarity via statistical models. In a system evaluation, we found that participants can effectively use \sys{} to generate UI automation scripts and complete tasks with high accuracy. We propose a generalizable neurosymbolic approach that combines the advantages of rule-based systems and neural networks. Our work can offer insights into future system and interaction designs that leverage semantic understanding in traditionally symbolic automation systems.

\begin{acks}
We thank all our participants and reviewers. This research was supported in part by the National Sciences and Engineering Research Council of Canada (NSERC) under Grant IRCPJ 545100 - 18, and the National Science Foundation under grant numbers CCF-2236233 and CCF-2123654.
\end{acks}
\\

\bibliographystyle{ACM-Reference-Format}
\bibliography{main}

\onecolumn

 
\appendix
\section{Appendix}
\vspace{-1pc}
\begin{figure*}[!h]
    \centering
    \includegraphics[width=0.73\linewidth]{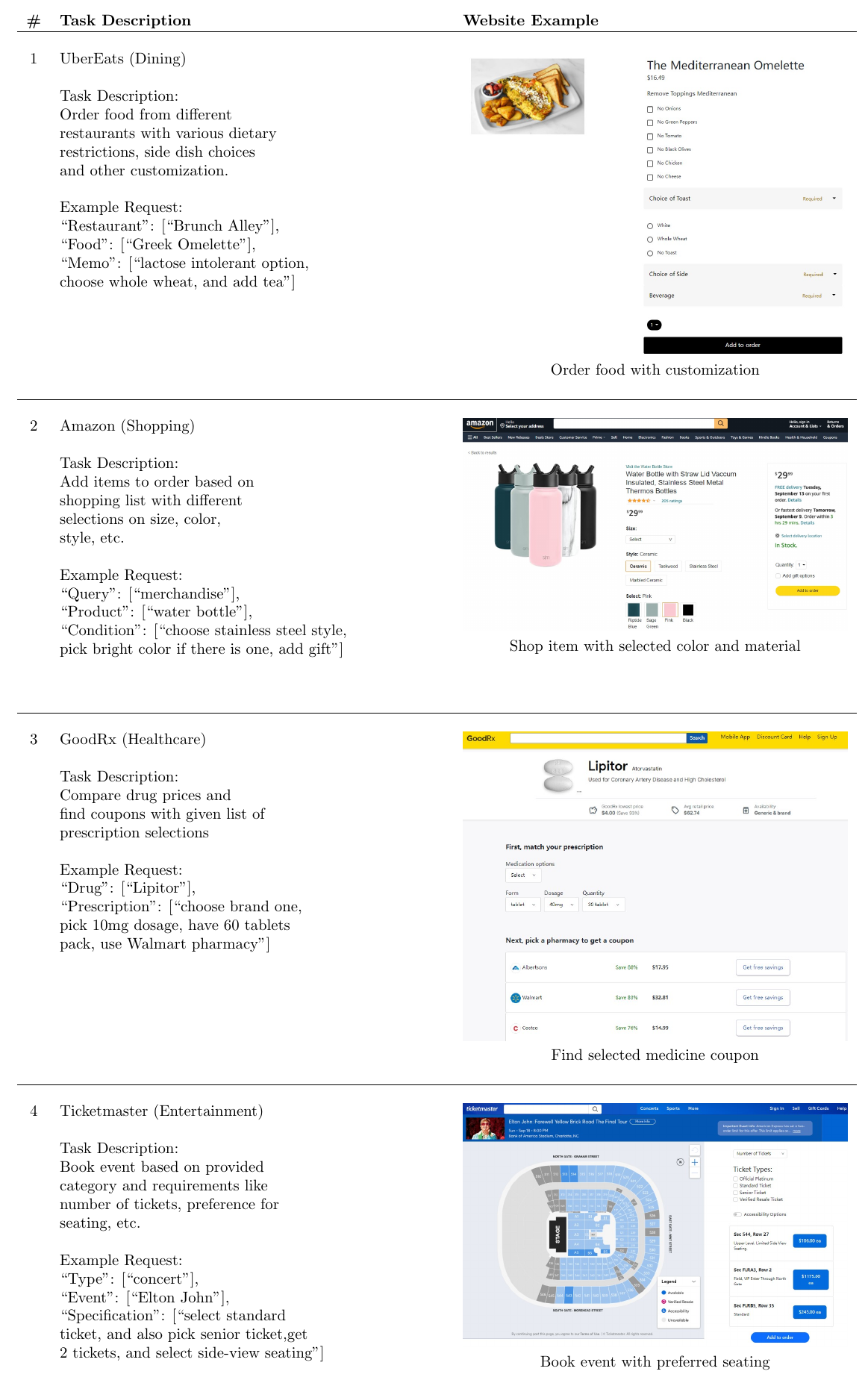}
    \caption{User study task descriptions with example web page UI.}
    \label{fig:task_table}
\end{figure*}

\end{document}

%% file: imports.tex
\usepackage{booktabs}   
\usepackage{subcaption} 

\usepackage{graphicx}
\usepackage{enumitem}
\usepackage[noend]{algpseudocode}
\usepackage{algorithmicx}
\usepackage[Algorithm]{algorithm}
\usepackage{bbm}
\usepackage{ragged2e}
\usepackage{float}
\usepackage{xcolor}

\usepackage{seqsplit}
\usepackage{hyperref}
\usepackage{url}
\usepackage{xspace}

\usepackage{color}
\usepackage{multirow}

\usepackage{array}
\usepackage{makecell}

\usepackage{balance}
\usepackage{lipsum}

\usepackage{rotating}
\usepackage{listings}
\usepackage{wrapfig}    

\algnewcommand\Input{\hspace{-12pt}\textbf{input: }}
\algnewcommand\Output{\hspace{-12pt}\textbf{output: }}
\algnewcommand\Invariant{\hspace{-12pt}\textbf{invariant: }}
\algrenewcommand\algorithmicindent{1em}

\newcolumntype{L}[1]{>{\raggedright\let\newline\\\arraybackslash\hspace{0pt}}m{#1}}
\newcolumntype{C}[1]{>{\centering\let\newline\\\arraybackslash\hspace{0pt}}m{#1}}
\newcolumntype{R}[1]{>{\raggedleft\let\newline\\\arraybackslash\hspace{0pt}}m{#1}}